\documentstyle[12pt,epsf,epsfig,a4wide]{article}
\begin{document}

\def\beq{\begin{equation}}
\def\eeq{\end{equation}}
\def\gtsim{\stackrel{>}{_\sim}}
\newcommand{\form}[1]{(\ref{#1})}
\newcommand{\gag}{g_{a\gamma\gamma}}

\begin{centering}
\begin{flushright}
arXiv: 0708.2646 [hep-ph] \\
CERN-TH/2007-146\\
August 2007
\end{flushright}

\vspace{0.1in}

{\Large {\bf Constraining axion by polarized prompt emission from gamma ray bursts}}

\vspace{0.4in}

%{\bf J.~Ellis}$^{a}$, {\bf N.E.~Mavromatos}$^{b}$,
%The Authors?
{\bf A.~Rubbia}$^{a}$ {\bf and A.S.~Sakharov$^{a,b}$}

\vspace{0.2in}

\vspace{0.4in}
 {\bf Abstract}

\end{centering}

\vspace{0.2in}

{\small \noindent 

\noindent A polarized gamma ray emission spread over a sufficiently wide energy
band from a strongly magnetized astrophysical object like gamma ray bursts
(GRBs) offers an opportunity to test the hypothesis of invisible axion. The
axionic induced dichroism of gamma rays at different energies should cause a
misalignment of the polarization plane for higher energy events relative to
that one for lower energies events resulting in the loss of statistics needed to
form a pattern of the polarization signal to be recognized in a detector.
According to this, any evidence of polarized gamma rays coming from an
object with extended magnetic field could be interpreted as a
constraint on the existence of the invisible axion for a certain parameter
range. Based on reports of polarized MeV emission detected in several GRBs
we derive a constraint on the axion-photon coupling. This constraint $\gag\le
2.2\cdot 10^{-11}\ {\rm GeV^{-1}}$ calculated for the axion mass
$m_a=10^{-3}~{\rm eV}$ is competitive with the sensitivity of CAST and becomes
even stronger for lower masses.

%Based on the report of polarized MeV emission from GRB021206 detected by
%RHESSI, we derive a new type of constraint on the axion-photon coupling. This
%constraint
%$\gag\le
%2.2\cdot 10^{-11}\ {\rm GeV^{-1}}$ calculated for the axion mass
%$m_a=10^{-3}~{\rm eV}$ is competitive with the sensitivity of CAST and becomes
%even 
%stronger for lower masses.
}

\vspace{0.5in}
\begin{flushleft}
CERN-TH/2007-146 \\
August 2007
\end{flushleft}

\vspace{0.25in}
\begin{flushleft}
$^a$ Swiss Institute of Technology, ETH-Z\"urich, 8093 Z\"urich, Switzerland \\
$^b$ TH Division, PH Department, CERN, 1211 Geneva 23, Switzerland \\ 
\end{flushleft}

%\end{titlepage}

\newpage

%\section{Introduction}
The Peccei-Quinn (PQ) mechanism~\cite{PQ} remains perhaps the most natural
solution to the CP problem in QCD. A new chiral $U_{PQ}(1)$ symmetry being
spontaneously broken at some large energy scale, $f_a$, and explicitly broken
by the color anomaly at QCD scale would allow for the dynamical vanishing of the
$\theta$ term and thus the restoration of the CP symmetry in strong
interactions. The pseudo-scalar field,  which drives the relaxation of
the $\theta$ term to zero is called axion. The most important phenomenological 
property of this axion is its two-photon vertex
interaction, which
allows for axion to photon conversion in the presence of an external electromagnetic and
magnetic fields~\cite{axion_gamma_vertex} through an interaction term
\beq
\label{gamma_vertex}
{\cal L}_{a\gamma}=-\frac{1}{4}\gag F_{\mu\nu}\tilde F^{\mu\nu}a=\gag{\bf
E\cdot B}a,
\eeq 
where $a$ is the axion field, $F$ is the electromagnetic field strength tensor,
$\tilde F$ its dual, ${\bf E}$ and ${\bf B}$ the electric and magnetic fields
respectively. The axion-photon coupling strength is quantified by 
\beq
\label{coupling}
\gag =\xi\frac{\alpha}{2\pi}\frac{1}{f_a},
\eeq
where $\alpha$ is the fine-structure constant and $\xi$ is an order
one~\footnote{Two quite distinct invisible axion models, namely the
KSVZ~\cite{ksvz} (hadronic axion) and the DFSZ~\cite{dfsz} one, lead to quite
similar $\gag$.} parameter which depends on the details of the electromagnetic
and color anomalies of the axial current associated with axion
field. For review of the properties of the invisible axion as well
as various types of constraints on the axion mass and couplings
see~\cite{kim,sikivie_rev}. 

According to~\cite{mimmo} the axion-photon mixing~\form{gamma_vertex} gives
rise to vacuum birefringence and dichroism, which are qualitatively identical
to those arising from the QED magnetized vacuum. If one
considers a beam of laying polarized monochromatic photons with frequency
$\omega$ and wave vector ${\bf k}$ propagating in a vacuo along in a
uniform magnetic field ${\bf B}$ laing at a nonvanishing angle $\phi$ with the
wave vector, due to the  birefringence the beam polarization becomes
elliptical at some distance from the source. On the other hand, the dichroism produces a
rotation of the ellipse's major axis with respect to the initial polarization.
In this letter we only consider the axionic dichroism induced rotation angle
$\epsilon$ of the polarization plane of an initially linearly polarized
monochromatic beam given in~\cite{mimmo,raffelt}:
\beq
\label{rotation}
\epsilon=N\frac{\gag^2B^2\omega^2}{m_a^2}
\sin^2\left(\frac{m_a^2L}{4\omega}\right)\sin 2\phi ,
\eeq
where $m_a$ is the mass of the axion, $L$ is the length of the magnetized
region, $N$ is the number of passes through the cavity with the magnetic field
if for instance a laser experiment like PVLAS~\cite{pvlas1,pvlas2} is considered. The validity of the approximation~\form{rotation} is provided if
the oscillation wavenumber 
\beq
\label{osc_num}
\Delta_{\rm osc}^2=\left(\frac{{m_a^2}-\omega_{\rm pl}}
{2\omega}\right)^2+B^2\gag^2
\eeq  
is dominated by the axion mass term. In fact,~\form{osc_num} pertains to the
situation in which the beam propagates in a magnetized
plasma, which gives rise to an effective photon mass set by the plasma
frequency $\omega_{\rm pl}=\sqrt{4\pi\alpha n_e/m_e}\simeq 3.7\cdot
10^{-11}\sqrt{n_e/{\rm cm}^{-3}}$~eV, where $n_e$ is the electron density and
$m_e$ is the electron mass. 
 Since the dichroism rotation, expected to arise in the QED vacuum,
and is suppressed with
respect to birefringence by factor $(B/B_{cr})^2$~\cite{adler}, where
$B_{cr}=m_e/e\simeq 4.4\cdot10^{13}$G, any observable rotation in vacuum should
be attributed to the axion-photon mixing term~\form{gamma_vertex}. 

 The polarization of the prompt gamma ray emission has been measured in four
bright GRBs: GRB021206, GRB930131, GRB960924 and GRB041219a. The first
measurements made in~\cite{boggs}  with Ranaty High Energy Solar Spectrometer
Imager (RHESSI) satellite~\cite{rhessi}, found a linear polarization,
$\Pi =(80\pm 20)\%$, of the gamma rays from GRB021206 across the spectral window
0.15-2~MeV. The analysis techniques have been challenged in~\cite{critic} and
defended in~\cite{defend}. Subsequent analyses made in~\cite{psi} confirmed the
results of~\cite{boggs} but at the lower level of significance. Later,
in~\cite{batsepol} the BATSE instrument on board of the Compton Gamma Ray
Observatory (CGRO)~\cite{cgro} has been used to measure, for two GRBs, the
angular distribution of gamma rays back-scattered by the rim of the Earth's
atmosphere: $35\%\le\Pi\le 100\%$ for GRB930131 and $50\%\le\Pi\le 100\%$ for
GRB960924. The analysing technique of~\cite{batsepol} is only sensitive to the
energy range 3-100~keV. Finally, the analysis~\cite{intpol} of GRB041219a
across the spectral window 100-350~keV has been performed using coincidence
events in the SPI (spectrometer on board of the INTEGRAL
satelite~\cite{integral}) and IBIS (the Imager on Board of the INTEGRAL
satelite). The polarization fraction of $\Pi =96^{+39}_{-40}\%$ was determined
 for this GRB. 
 
The above mentioned measurements are made using multiple events scattered
into a detector with geometry distinguishing capabilities (two adjacent
detectors in case of INTEGRAL or rotating detector in case of RHESSI). Because
Compton scatter angle depends on the polarization of incoming photons, the time
integrated polarization of the prompt gamma ray emission in a GRB can be of the
order of tens of per cent provided that the polarization angle does not vary
significantly during the whole duration of the GRB across the spectral range
analyzed.
For example the detected polarization signal in RHESSI arises from a
correlation between the time dependence of scattered photon flux and the angular
orientation of the satellite, which varies on a time scale similar to the burst
duration. The data~\cite{boggs_lv} indicate that a major contribution to the
flux comes from photons significantly distributed over at least the energy range
0.2-1.3~MeV. 

According to the 
Hillas~\cite{hillas}~\footnote{See Fig.~10. of~\cite{hillas_comp} for
the graphical compilation of~\cite{hillas}.}
diagram showing size and magnetic field strengths of different astrophysical
object the typical magnetic field in a GRB's engine can be estimated as
$B\simeq 10^9$~G over a region $L_{\rm GRB}\simeq
10^9$~cm. Therefore, one can observe that the constraint arises
from the fact that if the axionic dichroism induced angle of polarization
rotation~\form{rotation} in the given magnetic field were to differ by more then
$\pi /2$ over the energy range 0.2-1.3~MeV, as in the case of GRB021206, the
instantaneous polarization in the detector would fluctuate significantly for the
net time averaged polarization of the signal to be suppressed. Similar argument
can be applied to the polarization measurements based on INTEGRAL and BATSE
events.

To evaluate the bound we assume that in average the magnetic field is
sufficiently misaligned with the direction of the beam 
of gamma rays over the magnetized region implying that $\sin 2\phi\simeq 1$ and the
axion mass is large enough that for a given energy of gamma rays  $\omega$
\beq
\label{reg1}
\frac{m_a^2L_{GRB}}{4\omega}\gtsim\frac{\pi}{2}.
\eeq
 Therefore, the length
\beq
\label{l_pass}
L_{\rm pass}=\frac{2\pi\omega}{m_a^2}
\eeq 
can be interpreted as an analog of the passage length akin to that
one which defines the dimension of a magnetized cavity in a laser PVLAS 
experiment. Thus the number of passages in \form{rotation} is given by 
\beq
\label{N}
N=\frac{L_{\rm GRB}}{L_{\rm pass}}
\eeq
and thereby when the condition~\form{reg1} holds the accumulated polarization
rotation angle can be expressed from~\form{rotation} as
\beq
\label{rot1}
\epsilon =\frac{L_{\rm GRB}}{2\pi}\frac{\gag^2}{m_a^2}\omega B^2.
\eeq
Hence, the relative misalignment between the polarization planes of gamma
radiation at two different energies $\omega_1$ and  $\omega_2$ is given by
\beq
\label{relative}
\Delta\epsilon =\frac{L_{GRB}}{2\pi}\frac{\gag^2}{m_a^2}\Delta\omega B^2,
\eeq
where $\Delta\omega=|\omega_2-\omega_1|$.
The statistical pattern of the time integrated polarization signal from a
GRB in a detector is preserved for the energy range between $\omega_1$ and 
$\omega_2$ 
 provided that the relative
misalignment angle~\form{relative}  is less than $\pi /2$.  
This condition can be transformed into the bound on the axion-photon coupling as
\beq
\label{b1}
\gag\le\pi\frac{m_a}{B\sqrt{\Delta\omega L_{\rm GRB}}}.
\eeq 
\begin{figure} [t]
\begin{center}
\epsfig{file=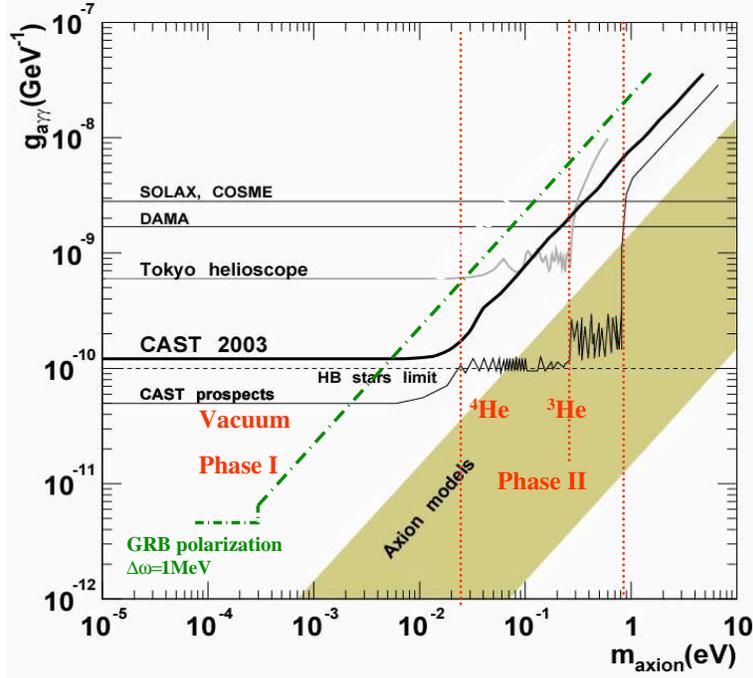,width=100mm,clip=}
\end{center}
\caption{\it The plot of the regions of $(m_a, \gag)$ space 
ruled out by various solar axion searches taken from~\cite{sikivie_rev} with
the bound of the present letter, estimated for the inner part of the energy
range 0.2-1.3~MeV applied for the polarization measurements of
GRB021206 (dashed
dotted line), superimposed.}
\label{castlimit}
\end{figure}
From the last inequality using $B=1.95\cdot 10^7~{\rm eV^2}$ and $\Delta\omega
L_{GRB} =5\cdot 10^{19}$ for the given magnetic field, energy difference $\Delta\omega =1$~MeV
and the length of magnetized region $L_{GRB}=10^9$~cm one obtains~\footnote{The
values are given in units $c=\hbar =1$, so that ${\rm 1T=195\ eV^2}$ and ${\rm
GeV\cdot cm=5\cdot 10^{13}}$.}
\beq
\label{gag1}
\gag\le 2.2\cdot 10^{-8}\frac{m_a}{1\ {\rm eV}}\ ({\rm GeV})^{-1},
\eeq 
where the inner part of the spectral window 0.2-1.3~MeV ($\Delta\omega\approx
1{\rm MeV}$)
reported in polarization analysis of GRB021206 has been used. 
However, for the axion mass
\beq
\label{m_crit}
m_a\le\sqrt{\frac{2\pi\omega}{L_{GRB}}}
\eeq 
the passage length~\form{l_pass} exceeds $L_{GRB}$ implying that the
rotation angle should be expressed as a constant:  
\beq
\label{rot2}
\epsilon =\frac{\gag^2}{16}(BL_{\rm GRB})^2.
\eeq
The upper edge $\omega_2\approx 1.3\ {\rm MeV}$ of the energy range considered,
together with  the condition~\form{m_crit}, defines the
critical mass
\beq
\label{m_crit1} 
m_{cr1}\approx 3.5\cdot 10^{-4}\ {\rm eV}. 
\eeq
Below this mass the rotation angle of the higher energy photons does not depend
either on their energy or the axion mass and is given by~\form{rot2}, while for
the lower energy edge of the polarized photons $\omega_1\approx 0.2 {\rm
MeV}$ the equation~\form{rot1} is still valid. Therefore, for the axion
mass bellow the 
critical one, $m_a\le m_{cr1}$, the polarization planes
misalignment angle
should be calculated as
\beq
\label{miss2}
\Delta\epsilon =B^2\gag^2L_{\rm GRB}\left(\frac{L_{\rm
GRB}}{16}-\frac{\omega_1}{2\pi
 m_a^2}\right).
\eeq 
 The expression~\form{miss2} 
holds to be positive down to the
mass 
\beq
\label{m_crit2}
m_{\rm cr2}= 4\sqrt{\frac{\omega_1}{2\pi L_{GRB}}}\approx 8\cdot 10^{-5}\ {\rm
eV}. 
\eeq
Requiring again that the misalignment angle~\form{miss2} does not exceed $\pi
/2$ in the axion mass range between $m_{\rm cr1}$ and $m_{\rm cr2}$ one
arrives to a bound, which can be well approximated by a constant
\beq
\label{lim_low_mass}
\gag\le \frac{2\sqrt{2\pi}}{BL_{\rm GRB}}\approx 5\cdot 10^{-12}\ ({\rm
GeV})^{-1}. 
\eeq
For the masses $m_a\ge 8\cdot 10^{-5}\ {\rm eV}$ the rotation
of the polarization plane is given by~\form{rot2} and misalignment does not
appear making the bound undefined for the axion mass bellow $m_{\rm cr2}$. 

In Fig.~1. we show the bounds~\form{gag1} and~\form{lim_low_mass}
superimposed on the recent results of CAST~\cite{cast} and other axion
helioscope experiments~\cite{rev_exp,Laz,Min,DAMA,Avi}. One can see that the
bound
obtained from the lack of a substantial  misalignment of polarization planes
of gamma
radiation at different energies within the energy range reported by polarization
 measurements of GRB021206 is competitive with the
current sensitivity of CAST for the axion masses below $10^{-3}$~eV. Moreover,
one
observes that it seems to be very unlikely that a laser experiment like PVLAS
could find a signal~\cite{pvlas1} corresponding 
to $\gag\simeq 10^{-5}\ {\rm GeV^{-1}}$ for the meV axions. Indeed, the lack of
the observation of the previously claimed rotation signal~\cite{pvlas1} has been
established recently by the PVLAS team after a substantial upgrade
of the facility has been made~\cite{pvlas2}. Also the PVLAS anomaly has not
found a support in the results of a pulsed "light shining through a wall"
experiment~\cite{wall}.  

Since according to~\cite{piran_pr} the electron number density in a GRB's
environment can be estimated as $n_e\simeq 10^{10}\ {\rm cm}^{-3}$ the
expression~\form{osc_num} is still axion mass dominated down to
$m_a\approx m_{\rm cr}$ for the given magnetic field, the energy of the gamma
radiation, $\omega\approx 1$~MeV, and constraints on $\gag$
calculated from~\form{gag1} and~\form{lim_low_mass} . Therefore, the validity of
the approximation~\form{rotation} holds in the range of the parameters
the bound in Fig.1. is imposed. The
minimal time scale of variability of GRBs light curves is estimated
to be about 0.1 sec~\footnote{See, for example, the analysis in~\cite{wave}.}.
This implies that the typical extention of the GRB's engine is indeed
compatible with $L_{GRB}\approx 10^9$~cm we used for the evaluation of the
bound. Conservation of magnetic field energy at the rest wind frame of fireball
shell model of the GRB's engine~\cite{piran_pr} implies at any radial distance
$r$, in the fireball environment, $4\pi r_0^2B_0^2=4\pi r^2B^2$, leading to the
relation $B=B_0(r_0/r)$, where $B_0$ and $r_0$ are the magnetic field strength
and the size of the central part of the fireball. Typically the central
part of the fireball can be represented  by a neutron star of radius $r_0\approx
10^6$~cm with magnetic field of $B_0\approx 10^{12}$~G. Therefore the
strength of the magnetic field at the distance $r=L_{GRB}$ corresponds to 
$B\approx 10^9$~G, which is in a good agreement
with the values taken from~\cite{hillas,hillas_comp}.

The limit obtained becomes by factor
$\sqrt{1{\rm MeV}/\Delta\omega_{I,B}}$ weaker if we apply the width
$\Delta\omega_{I}\approx 250$~keV of the energy bands for GRB041219a 
 detected by INTEGRAL or $\Delta\omega_{B}\approx 100\ {\rm keV}$\- for
GRB930131 and
GRB960924 detected by BATSE. This implies that $\gag\le
4.4\cdot 10^{-11}\ {\rm GeV^{-1}}$ and $\gag\le
6.9\cdot 10^{-11}\ {\rm GeV^{-1}}$ for INTEGRAL and BATSE measurements
 respectively calculated for the axion mass
$m_a=10^{-3}~{\rm eV}$. A stronger
constraint could be obtained from the GRBs considered by taking into account
more precisely the spectral characteristics of the signal and statistical
criteria for loosing the polarization pattern in the detectors.   Of course, the
robustness
should find its confirmation in further detection of gamma polarized signals
from other GRBs in the similar energy range. In principle,
apart from the mentioned instruments SWIFT~\cite{swift} satellite is also
capable of polarimetry~\cite{pol_hep} in 300~keV-10~MeV energy band.  

In recent years, numerous efforts have been initiated to develop instruments
with the sensitivity required for astrophysical polarimetry over 100~eV to
10~GeV band~\cite{pol_astro}. Time projection chambers (TPCs), with their
high-resolution event imaging capability, are an integral part of some of this
efforts. At the energy band 300~keV-10~MeV, Compton polarimeters based on the
use of high-Z scattering elements (coupled with high-Z absorbers) become
viable. For example, the Ge double scatter approach used by RHESSI becomes most
effective at energies above 300~keV. Liquid rare-gas TPCs are being pursued as
large effective area Compton telescopes, while electron-tracking gas TPCs are a
component of Compton telescopes with the lowest background, especially for the
most polarization sensitive events. In the pair production regime, which is
effective in the energy range 2~MeV-10~GeV, only TPCs currently offer the
hope~\cite{pol_astro} of tracking electron-positron pairs or recoil electrons
with the accuracy and efficiency required for astronomical polarimetry.

\end{document}